\pdfoutput=1
\newcommand{\Title}{
Learning heterogeneous delays in a layer of spiking neurons for fast motion detection
}
\newcommand{\ShortTitle}{
Learning heterogeneous delays of spiking neurons for motion detection
}
\newcommand{\FirstAG}{Antoine}
\newcommand{\LastAG}{Grimaldi}

\newcommand{\FirstLP}{Laurent U}
\newcommand{\LastLP}{Perrinet}

\newcommand{\EmailLP}{laurent.perrinet@univ-amu.fr}
\newcommand{\EmailAG}{antoine.grimaldi@univ-amu.fr}

\newcommand{\Department}{Institut de Neurosciences de la Timone}
\newcommand{\Affiliation}{Aix Marseille Univ, CNRS}%
\newcommand{\Street}{27 boulevard Jean Moulin}%
\newcommand{\PostCode}{13005}%
\newcommand{\City}{Marseille}%
\newcommand{\Country}{France}%
\newcommand{\Keywords}{time code, event-based computations, spiking neural networks, motion detection, efficient coding, logistic regression
}
\newcommand{\Funding}{
This research was funded by the European Union ERA-NET CHIST-ERA 2018 research and innovation program under grant agreement ANR-19-CHR3-0008-03 (``\href{APROVIS3D}{http://aprovis3d.eu/}''). 
LP received funding from the ANR project ANR-20-CE23-0021 (``\href{AgileNeuroBot}{https://laurentperrinet.github.io/grant/anr-anr/}'')
and from A*Midex grant number AMX-21-RID-025
(``\href{Polychronies}{https://laurentperrinet.github.io/grant/polychronies/}''). 
}
\newcommand{\Acknowledgments}{
Thanks to 
Salvatore Giancani, Hugo Ladret, Camille Besnainou, Jean-Nicolas Jérémie, Miles Keating, and Adrien Fois for useful discussions during the elaboration of this work. 
\Funding %
A CC-BY public copyright license has been applied by the authors to the present document and will be applied to all subsequent versions up to the Author Accepted Manuscript arising from this submission, in accordance with the grant’s open access conditions. 
}
\newcommand{\DataAvailability}{
This works is made reproducible. The code reproducing the manuscript and all figures is available on~\href{https://github.com/SpikeAI/2023_GrimaldiPerrinet_HeterogeneousDelaySNN}{GitHub}. It also contains supplementary figures and results. Find also the associated~\href{https://www.zotero.org/groups/4776796/fastmotiondetection}{zotero group} used to gather relevant literature on the subject.
}
\documentclass[default]{sn-jnl}

\usepackage[T1]{fontenc}
\usepackage[utf8]{inputenc}
\jyear{2023}%
\theoremstyle{thmstyleone}%
\theoremstyle{thmstyletwo}%

\theoremstyle{thmstylethree}%

\newcommand{\seeFig}[1]{see Fig.~\ref{fig:#1}}
\usepackage{amsmath,amssymb,amsfonts} 
\usepackage{physics} 
\usepackage{siunitx}

\newcommand{\ms}{\si{\milli\second}}%
\newcommand{\presynaddr}{a} 
\newcommand{\postsynaddr}{b} 
\newcommand{\numevent}{N_{ev}} 
\newcommand{\presynaddrspace}{\mathcal{A}} 
\newcommand{\postsynaddrspace}{\mathcal{B}} 
\newcommand{\Npol}{N_\text{p}} 
\newcommand{\arank}{r} 
\newcommand{\bias}{\beta} 
\newcommand{\synapse}{\mathcal{S}} 
\newcommand{\synapticweight}{w} 
\newcommand{\synapticdelay}{\delta} 
\newcommand{\ranksyn}{s} 
\newcommand{\Nsyn}{N_{s}} 
 
\newcommand{\timev}{t} 
\newcommand{\polev}{p} 
\newcommand{\event}{\epsilon} 
\newcommand{\Nx}{N_\text{X}}
\newcommand{\Ny}{N_\text{Y}}
\newcommand{\Ntime}{N_\text{T}}
\newcommand{\kernel}{K} 
\newcommand{\Kx}{K_\text{X}}
\newcommand{\Ky}{K_\text{Y}}
\newcommand{\Ktime}{K_\text{T}}

\newcommand{\class}{c} 
\newcommand{\Nclass}{N_\class} 

\usepackage{soulutf8}
\usepackage{color}

\begin{document}
\title[\ShortTitle]{\Title}
\artnote{This article is published as part of the Special Issue on "What can Computer Vision learn from Visual Neuroscience?"}
\author{\fnm{\FirstAG} \sur{\LastAG}}\email{\EmailAG}
\author{\fnm{\FirstLP} \sur{\LastLP}}\email{\EmailLP}
\affil{\orgdiv{\Department}, \orgname{\Affiliation}\\ \orgaddress{\street{\Street}, \city{\City}, \postcode{\PostCode}, \country{\Country}}}
\abstract{
The precise timing of spikes emitted by neurons plays a crucial role in shaping the response of efferent biological neurons. This temporal dimension of neural activity holds significant importance in understanding information processing in neurobiology, especially for the performance of neuromorphic hardware, such as event-based cameras. Nonetheless, many artificial neural models disregard this critical temporal dimension of neural activity. In this study, we present a model designed to efficiently detect temporal spiking motifs using a layer of spiking neurons equipped with heterogeneous synaptic delays. Our model capitalizes on the diverse synaptic delays present on the dendritic tree, enabling specific arrangements of temporally precise synaptic inputs to synchronize upon reaching the basal dendritic tree. We formalize this process as a time-invariant logistic regression, which can be trained using labeled data. To demonstrate its practical efficacy, we apply the model to naturalistic videos transformed into event streams, simulating the output of the biological retina or event-based cameras. To evaluate the robustness of the model in detecting visual motion, we conduct experiments by selectively pruning weights and demonstrate that the model remains efficient even under significantly reduced workloads. In conclusion, by providing a comprehensive, event-driven computational building block, the incorporation of heterogeneous delays has the potential to greatly improve the performance of future spiking neural network algorithms, particularly in the context of neuromorphic chips.
}

\maketitle
\section{Introduction}
\label{sec:intro}
The human brain has the remarkable capacity to react efficiently at any given time while consuming a reasonable amount of energy, in the order of $20$ watts. This system is the result of millions of years of natural selection, and a striking difference with artificial neural networks is the representation that both use. Indeed, our computers use digital values and for instance, the convolutional neural networks (CNNs) which are used for processing images represent the flow of information from one layer to another using tensors of floats. These networks store visual information densely across the visual topography, with different translation-invariant properties represented in different channels. CNNs have achieved state-of-the-art performance for some computer vision tasks, such as image recognition. These networks are also known to mimic several properties of the biological visual system, such that each can be assigned a ``brain score''~\citep{schrimpf_brain-score_2020}. However, this score does not take into account key aspects of the efficiency of biological systems, such as inference speed (usually several times the biological time) or the energy consumption of this mesocopic model of the brain, which is about $360$ watts on a standard GPU (NVIDIA RTX 3090). In the vast majority of biological neural networks, on the other hand, information is represented as \emph{spikes}, prototypical all-or-nothing (binary) events whose only parameters are their timing and the address of the neuron that fired the spike~\citep{paugam-moisy_computing_2012}. Spiking neural networks (SNNs), known as the third generation of artificial neural networks, incorporate this temporal dimension into the way they perform their computations. One example of a SNN is the SpikeNet algorithm, which takes a purely temporal approach by encoding information using at most one spike per neuron~\citep{delorme_spikenet_1999}. Alternatively, the SNN implemented in the SpikeProp algorithm~\citep{bohte_error-backpropagation_2002} that uses the exact timing of spikes and learns the structure of the network by minimizing a specifically defined cost function. This was recently extended using the surrogate gradient method which is now widely used in attempts to transfer the performance of CNNs to SNNs~\citep{zenke_remarkable_2021}. However, the performance of SNNs still lags behind that of networks using an analog, firing rate-based representation. The question of the advantage of using spikes in machine learning and computer vision remains open.

In a recent review, we reported previous theoretical and  experimental evidence for the use of precise spiking motifs in biological neural networks~\cite{grimaldi_precise_2023}. In particular, \citet{abeles_role_1982} asked whether the role of cortical neurons is the integration of synaptic inputs or rather the detection of coincidences in temporal spiking motifs. While the first hypothesis favors the firing rate coding theory, the second emphasizes on the importance of temporal precision in neural activity. Since then, numerous studies have demonstrated the emergence of synchronous activity within a neuronal population~\citep{riehle_spike_1997, davis_spontaneous_2021}, efficient coding using precise spike timings~\citep{Perrinet2002,perrinet_coding_2004, gollisch_rapid_2008}, or precise timing in the auditory system~\citep{deweese_binary_2002, carr_circuit_1990}. All these findings, and more~\citep{bohte_evidence_2004}, highlight the importance of the temporal aspect of the neural code and suggest the existence of spatiotemporal spiking motifs in biological spike trains. In neural models, the definition of heterogeneous delays~\citep{guise_bayesian_2014, zhang_supervised_2020, nadafian_bio-plausible_2020} allows the efficient detection of these spatiotemporal motifs embedded in the spike train. Such spatiotemporal motifs present in neural activity may form useful representations to perform computations for various cognitive tasks using the synchrony of spikes reaching the soma of a neuron. In particular,~\citet{izhikevich_polychronization_2006} introduced the notion of the polychronous group as a spiking motif defined by a subset of neurons with different, but precise, relative spiking delays. This delay between a pair of connected neurons is defined as the time between the emission of a spike at the soma of the afferent neuron and its arrival at the soma of the efferent neuron. Importantly, due to the variety of weights and delays within a population, representations using polychronous groups have, in theory, a much higher information capacity than a firing rate-based approach.

The present paper proposes a real-world application that extends a recently proposed model of spiking neurons with heterogeneous synaptic delays~\cite{grimaldi_learning_2022}. This model was trained to solve a simplified motion detection task on a synthetic event-based dataset generated by moving parameterized textures, and provides a first demonstration that formal neurons can exploit the precise timing of spikes to detect motion thanks to heterogeneous delays. Here, we extend these results to a much more complex and natural setting. First, we define the ecological cognitive task that the model must solve with the different datasets on which it will be tested. Instead of the synthetic textures used previously, we use natural scenes synthesized from natural images translated by biologically inspired saccadic movements. We then develop the Heterogeneous Delays Spiking Neural Network (HD-SNN) model from efficiency principles and derive a learning rule to adapt the weights of each heterogeneous synaptic delay using gradient descent. Applied to this detection task, we study the emergence of spatiotemporal spiking motifs when this single layer of spiking neurons is trained in a supervised manner. We investigate the efficiency of the motion detection mechanism and, in particular, its resilience to synaptic weight pruning. Indeed, once trained, the amount of event-driven computation could be drastically reduced by removing weak synapses while maintaining peak performance for the classification task. In this way, we will be able to show how such a model can provide an efficient solution to the energy/accuracy tradeoff.
%
\section{Methods}
\label{sec:methods}
This paper aims to investigate the capability of the HD-SNN model to effectively learn and solve a motion detection task using realistic event-based data streams, as typically captured by event-based cameras, also known as Dynamic Vision Sensors (DVS). DVS are designed to mimic the signal transmitted from retinal ganglion cells to the visual cortex through the optic nerve. The events in these data streams are binary in nature and should provide sufficient information for performing fast and efficient motion detection. In this study, we first outline the task definition, specifying the requirements for motion detection. Subsequently, we describe the HD-SNN model employed for inferring motion, highlighting its key characteristics and architecture. Lastly, we elaborate on the training procedure employed to train the HD-SNN model for the specific motion detection task.
%
\subsection{Task definition: motion detection in a synthetic naturalistic event stream}\label{sec:task}
\begin{figure}%
    \centering
    \includegraphics[width=0.95\linewidth]{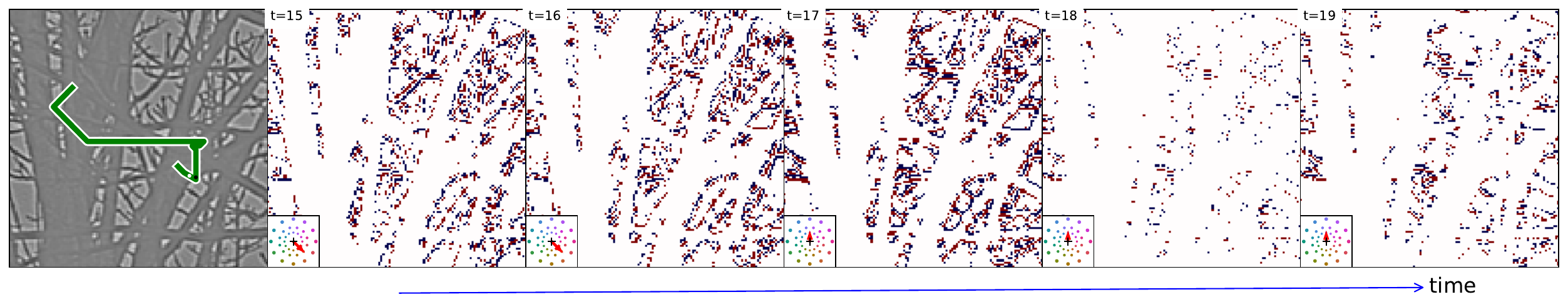}
    \caption{
        {\bf Motion Detection Task.} To generate realistic event-based dynamic scenes, we mimic the effect of minute saccadic eye movements on a large natural scene ($1024\times1024$) by extracting an image ($128\times128$) which center is moving dynamically according to a jagged random walk. \textit{(Left)}~We show an instance of this trajectory (with a length of $200~\ms$, green line) superimposed on the luminance contrasts observed at time step $t=15~\ms$. \textit{(Right)}~The dynamics of this image, translated according to the saccadic trajectory, produces a naturalistic movie, which is then transformed into an event-based representation. We show snapshots of the resulting synthetic event stream at different time steps (from $t=15~\ms$ to $t=19~\ms$, these frames are marked on the trajectory by a white and black dot, respectively, in the left inset). Mimicking the response of ganglion cells in the retina, this representation encodes at each pixel all-or-none increases or decreases in luminance, i.e., ON (red) and OFF (blue) spikes. In the lower left corner of the snapshots, we show the corresponding instantaneous motion vector (red arrow). Note the change in the direction of motion between the third and fourth frames, and also that contours parallel to the motion produce fewer luminance changes, the so-called aperture problem, and thus relatively fewer spikes.
        }
    \label{fig:motion_task}
\end{figure}

To train and validate our model which uses supervised learning, we need to define a visual dataset for which we explicitly know the ground truth motion, that is, direction and speed. To achieve this, we define a procedure for animating a natural visual scene with virtual eye movements, similar to those used in  studies from neurobiology~\citep{vinje_sparse_2000, baudot_animation_2013} or computational neuroscience~\citep{kremkow_push-pull_2016}. First, we draw a trajectory inspired by biological eye movements. Indeed, these movements allow us to dynamically actuate the center of vision in the visual field. In animals with a fovea, this is particularly useful as it allows the area with the highest density of photoreceptors to be moved, for example, to a point of interest in the environment. A specific mechanism for this function are saccades, which are rapid eye movements that reposition the center of vision. In humans, saccades are very frequent (on average $2$ per second~\citep{dandekar_neural_2012}), very fast (about $80~\ms$), and have a wide range of speeds. On a more microscopic scale, the human gaze moves with minute micro-saccades which trajectories are similar to a Brownian trajectory~\citep{poletti_head-eye_2015}. To preserve the full generality of the task, we define eye movements using such a form of random walk~\citep{engbert_integrated_2011}: We first define a finite set of possible motions in polar coordinates as the Cartesian product of the $12$ regularly spaced directions and $3$ geometrically spaced speeds. Next, we define a saccadic path as a sequence of time segments whose durations are drawn from a Poisson distribution with a mean block length of $24~\ms$, similar to a Lévy flight~\citep[p. 289]{mandelbrot_fractal_1982}.  Finally, assuming motion is stationary during saccades, and that motions for each flight are drawn uniformly and independently, the global trajectory is generated by integrating this motion sequence. This generative model yields trajectories that are qualitatively very similar to those observed for human eye movements (\seeFig{motion_task}-\textit{(Left)}).

Once these eye trajectories are generated, we can apply them to a visual scene. For this purpose, we selected a database of $100$ large-scale natural images that were previously used to study the statistics of natural images~\citep{vanHateren1998}. Note that these were pre-processed to be in grayscale and to equalize (i.e., whiten) the energy in each frequency band, similarly to a process known to occur in the retino-thalamic pathway~\citep{dan_efficient_1996}. The full-scale images are $1024 \times 1024$ in size, and we crop images of size $128 \times 128$ positioned around the center of gaze at each time step. We discretize time in $1~\ms$ bins and produce movies of duration $N_T = 200~\ms$. To avoid boundary effects, we randomly position the full trajectory in image space so that the sub-image is translated using the position given by the trajectory at each time step and without touching the boundaries. At each time step, the translation is computed using a coordinate roll in the horizontal and vertical dimensions, followed by a sub-pixel translation defined in Fourier space~\citep{perrinet_sparse_2015}. Note that the magnitude of the displacement is relative to the time bin, and we have defined the unit speed to correspond to a movement of one pixel per frame (i.e., per time bin of $1~\ms$).

To transform each movie into events, as recorded by a DVS, we compute a residual gradient image which we initialize at zero. We then compute the temporal gradient of the pixels' intensity over two successive frames. For a given pixel and time stamp, an event is generated when the absolute value of this gradient exceeds a threshold. The event has either an OFF or ON polarity, depending on whether the gradient is negative or positive. The signed threshold is then subtracted from the residual gradient image. When applied to the whole movie, the event stream is similar to the output of a neuromorphic camera~\citep{Gallego2022}, i.e. a list of events defined by $x_\arank$ and $y_\arank$ (their position on the pixel grid), their polarity $\polev_\arank$ (ON or OFF), and their time $\timev_\arank$ (\seeFig{motion_task}-\textit{(Right)}). Ultimately, the goal of the model is to infer the correct motion solely by observing these events.
\subsection{The HD-SNN model}
%
\begin{figure}
    \centering
    \includegraphics[width=0.980\linewidth]{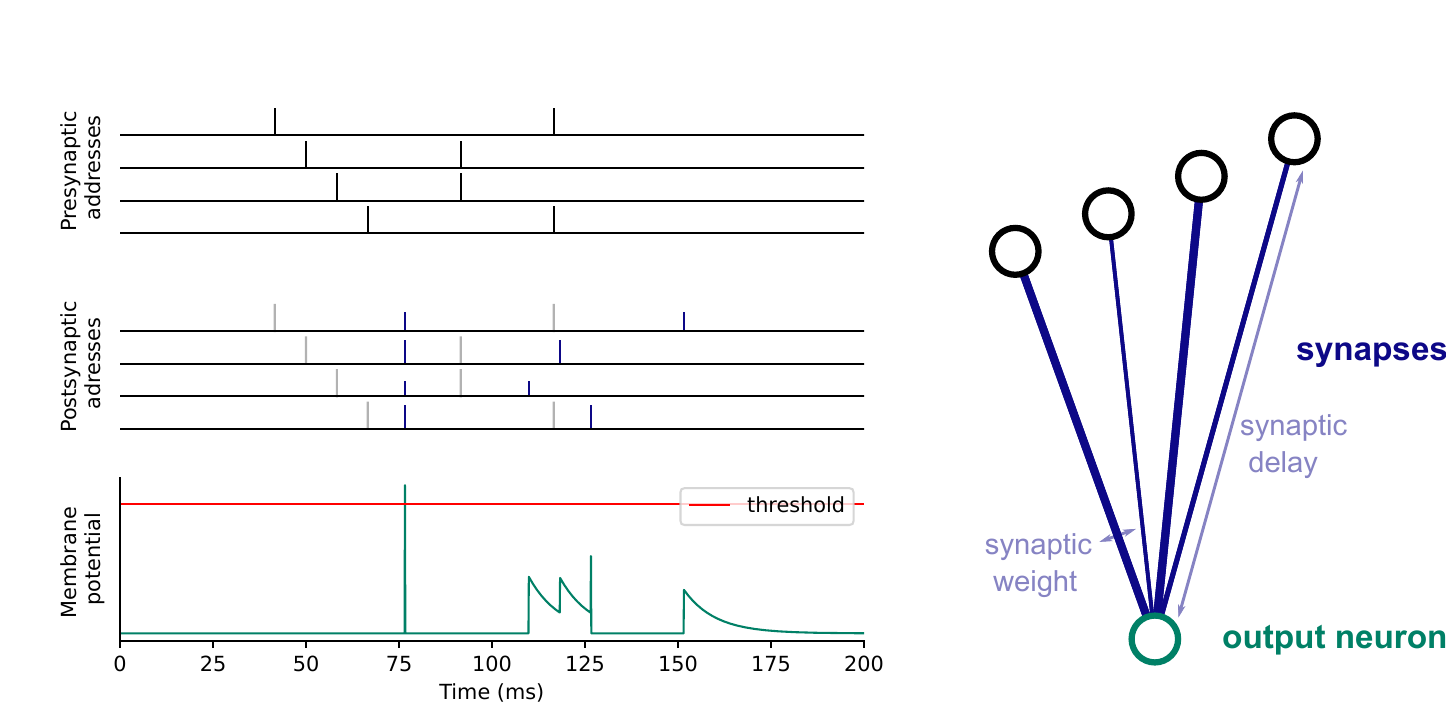}
    \caption{\textbf{Core mechanism of the HD-SNN model.} \textit{(Left-Top)}~Four presynaptic neurons show some spiking activity in which a spiking motif is embedded (starting at time $t=50~\ms$). \textit{(Right)}~An illustration of a spiking neuron with different synaptic weights (represented by the thickness of the synapses) and different synaptic delays (represented by the length of the synapses). \textit{(Left-Middle)}~Each spike is weighted by the synaptic weights (height of the blue bars) and shifted in time according to the synaptic delays on each respective synapse (input spikes are shown in light gray for comparison). As a result, the spikes from the spiking motif are synchronized as they reach the soma of the postsynaptic neuron. \textit{(Left-Bottom)}~These spikes are then integrated, and contribute to a modification of the membrane potential of the output neuron according to the neural activation function. In this example, we use the activation function of a Leaky Integrate-and-Fire neuron. The first spiking motif is synchronized by the synaptic delays and causes a sudden rise in the membrane potential of the postsynaptic neuron. An output spike is emitted at time $t=75~\ms$ when the membrane potential reaches the threshold, and it is then reset.}%
    \label{fig:izhikevich}%
\end{figure}%
%
In this task, the input consists of a stream of events or \emph{spikes}, a representation common to the signal obtained from an event-based camera or from a neurobiological recording of the activity of single units. Formally, this can be defined as a list of tuples, each tuple representing the neural address and timestamp. We denote this list as $\event = \{(\presynaddr_\arank, \timev_\arank)\}_{\arank \in [1,\numevent]}$ where $\numevent \in \mathbb{N}$ is the total number of events in the data stream and the rank $\arank$ is the index of each event in the list of events (\seeFig{izhikevich}-\textit{Left-Top} for an illustration). Each event has a time of occurrence $\timev_\arank$ and an associated address $\presynaddr_\arank$. Events are usually ordered by their time of occurrence. We define the address space $\presynaddrspace$, which consists of the set of possible addresses. In neurobiological spiking activity, this may be the identified set of recorded neurons. For neuromorphic hardware like the output of a DVS or our task, this can be defined as $[1, \Npol] \times [1, \Nx]\times[1, \Ny] \subset \mathbb{N}^3$, where $\Npol$ is the number of polarities ($\Npol=2$ for the ON and OFF polarities encoded in event-based cameras) and $(\Nx, \Ny)$ is the height and width of the image in pixels. Thus, each address $\presynaddr_\arank$ is typically in the form $(\polev_\arank, x_\arank, y_\arank)$ for event-based cameras.

In the HD-SNN model, neurons $\postsynaddr \in \postsynaddrspace$ are connected to presynaptic afferent neurons from $\presynaddrspace$ using realistic synapses. In biology, a single cortical neuron typically has several thousand synapses. Each synapse can be defined by its synaptic weight and its delay, that is, the time it takes for a spike to travel from the soma of the presynaptic neuron to the soma of the postsynaptic neuron. Note that a neuron can contact another afferent neuron with different delays through different synaptic connections. By scanning all postsynaptic neurons $\postsynaddr$, we may thus define the full set of $\Nsyn$ synapses, as $\synapse = \{(\presynaddr_\ranksyn, \postsynaddr_\ranksyn, \synapticweight_\ranksyn, \synapticdelay_\ranksyn)\}_{\ranksyn \in [1,\Nsyn]}$, where each synapse is associated with a presynaptic address $\presynaddr_\ranksyn$, a postsynaptic address $\postsynaddr_\ranksyn$, a weight $\synapticweight_\ranksyn$, and a delay $\synapticdelay_\ranksyn$. This defines the full connectivity of the HD-SNN model (\seeFig{izhikevich}-\textit{Right} for an illustration of the connectivity of one neuron with synaptic weights and delays). 

Of interest is to define the emitting field of a presynaptic neuron $\synapse_\presynaddr = \{(\presynaddr_\ranksyn, \postsynaddr_\ranksyn, \synapticweight_\ranksyn, \synapticdelay_\ranksyn) \| \presynaddr_\ranksyn=\presynaddr\}_{\ranksyn \in [1,\Nsyn]} \subset \synapse$, or also the receptive field of a postsynaptic neuron $\synapse^\postsynaddr = \{(\presynaddr_\ranksyn, \postsynaddr_\ranksyn, \synapticweight_\ranksyn, \synapticdelay_\ranksyn) \| \postsynaddr_\ranksyn=\postsynaddr\}_{\ranksyn \in [1,\Nsyn]}  \subset \synapse$. In particular, when driven by a stream of spikes $\event = \{(\presynaddr_\arank, \timev_\arank)\}_{\arank \in [1,\numevent]}$, each incoming spike is multiplexed by the synapses of the receptive field $\synapse^\postsynaddr$ of postsynaptic neuron $\postsynaddr$. This results in a weighted event stream (\seeFig{izhikevich}-\textit{Left-Middle}) for each postsynaptic neuron $\postsynaddr$: 
%
\begin{equation}\label{eq:stream_b}
\event_\postsynaddr = \{(\presynaddr_\arank, \synapticweight_\arank, \timev_\arank+\synapticdelay_\ranksyn) \| \presynaddr_\arank = \presynaddr_\ranksyn \}_{\arank \in [1,\numevent], \ranksyn \in \synapse^\postsynaddr}
\end{equation}
%
In biology, this new stream of events is naturally ordered in time as events reach the soma of postsynaptic neurons. In simulations, however, it should be properly reordered. Once transformed by the synaptic connectivity, this weighted event stream may be integrated, for instance as the membrane potential of a Leaky-Integrate-and-Fire neuron (\seeFig{izhikevich}-\textit{Left-Bottom}), yet the activation function of the HD-SNN neurons can be selected from the full range of spiking neuron response functions. Importantly, this activation function has to be such that when postsynaptic neurons are activated at their soma by a specific spatiotemporal motif imprinted in the synaptic set, and such that these spikes converge at the soma in a synchronous manner, the discharge probability should increase. In this subsection, we have briefly defined the HD-SNN model in all generality (see~\citep{KeatingPerrinet23ICANN} for a more specific description and treatment), and in the next subsection we describe an implementation of our model adapted to the motion detection task. 
%
\subsection{Application of HD-SNN to motion detection}
In fact, it is possible to adapt the HD-SNN model specifically for common computer vision tasks. First, neural addresses are defined to represent the range of possible positions and polarities. Second, to simulate such event-based computations on standard CPU- or GPU-based computers and to benefit from parallel computing acceleration, we transform the temporal event-based representation into a dense dicretized representation. Indeed, by using this discretization, we transform any event-based input from an event-based camera into a Boolean matrix $A \in \{0, 1 \}^{\Npol \times \Ntime \times \Nx \times \Ny}$ defined for all polarities $p$, times $t$, and space coordinates $x$ and $y$. The values are, by definition, equal to zero, except when events occur: $\forall \arank \in [1,\numevent]$, $A(\polev_\arank, t_\arank, x_\arank, y_\arank)=1$. Similarly, one may discretize the connectivity of the HD-SNN model defined above. The longest synaptic delay defines the depth $\Ktime$ of the kernel, so that all possible delays associated with the different presynaptic addresses are represented. In particular, for each class $\class$ of the supervision task, the entire synaptic set can be represented as a kernel, which is represented by the dense matrix $\kernel$ of size $(\Nclass, \Npol, \Ktime, \Kx, \Ky)$, where $\Nclass$ is the number of classes, $\Ktime$ the number of delays and $\Kx$ and $\Ky$ are the number of pixels in each spatial dimension. To keep the analogy with the HD-SNN model, $\kernel$ gives the synaptic set that defines the weight of all synapses $\ranksyn$ defined as a function of their class $\class$, polarity, synaptic delay and relative position: $\forall {\polev \in [1, \Npol], \delta_\timev \in [1, \Ktime], \delta_x \in [1, \Kx], \delta_y \in [1, \Ky]}, \; \kernel(\class, \polev, \synapticdelay_\timev, \delta_x, \delta_y) = \synapticweight_\ranksyn$. In our simulations, we define as many classes as the number of motions (directions and velocities): $\Nclass = 12 \times 3$ and set the size of the model's kernel to $(\Nclass, \Npol, \Ktime, \Kx, \Ky) = (36, 2, 21, 17, 17)$. Such a kernel defines a dense representation of the full synaptic set. 

\begin{figure}
    \centering
    \includegraphics[width=\linewidth]{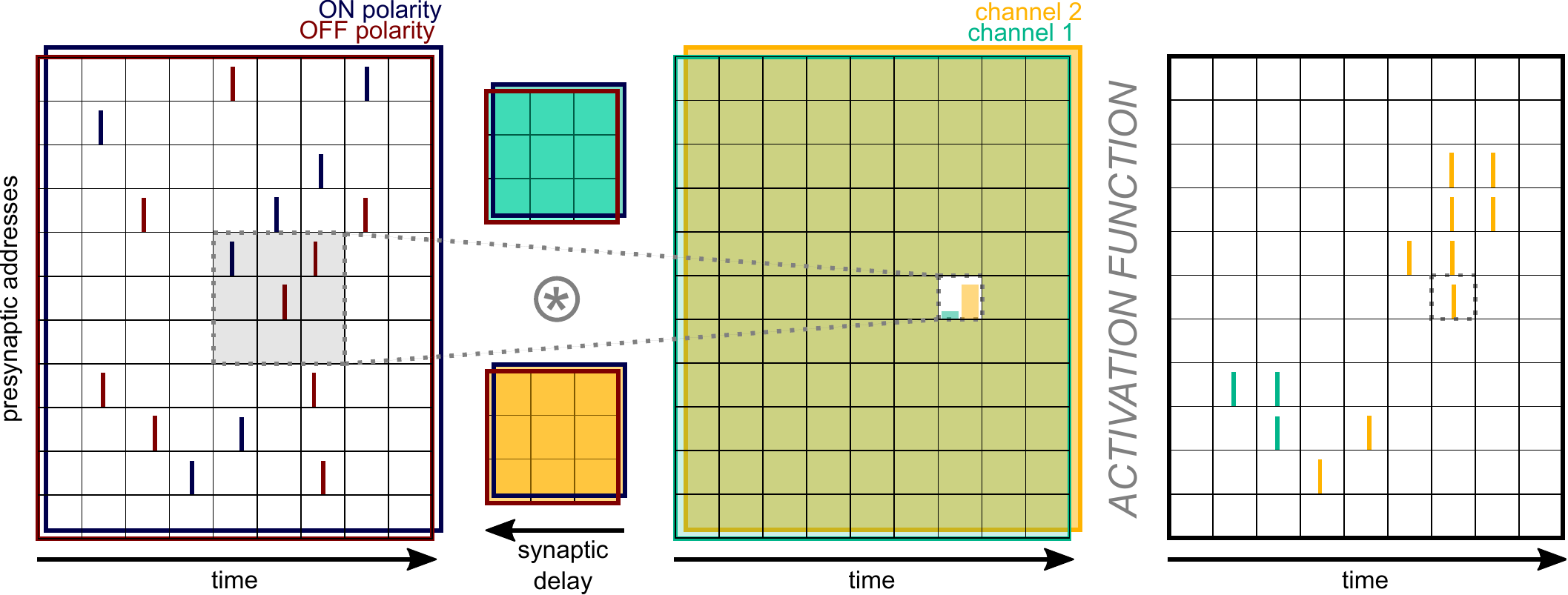}
    \caption{
        \textbf{Applying HD-SNN to the task of motion detection.} \textit{(Left)}~We plot a 2D representation of the input event stream as a raster plot (showing ON spikes in red and OFF spikes in blue for each presynaptic address and time). A spatiotemporal convolution is applied to the dense representation of the input with $2$ different convolution kernels (the green and orange kernels), which define the output channels. The convolution is summed over the two polarities. Since we have two axes $X$ and $Y$ to represent the presynaptic addresses, like the pixel grid of a DVS, this results in a 3D convolution. Here we simplify the illustration to a 2D representation and to 2 possible classes (green and orange) associated to two different directions of motion. \textit{(Middle)}~For each position (address, time) one can compute the activation resulting from the convolution. The output of the convolution is processed by the nonlinearity of the MLR model (i.e., the sigmoid function). The output of the MLR gives a probability for each class associated with a particular kernel (colored bars in the highlighted pixel). \textit{(Right)}~By adding a spiking mechanism, here a winner-takes-all associated with thresholding, we obtain as output of the HD-SNN model a new spike train with the different spikes associated with a particular motion class. Note that the position of the output spikes does not systematically correspond to the position of the input spikes, but only when enough evidence is reached.
    }
    \label{fig:model}
\end{figure}

Then, it can be noted that by using a discretization, the computational block used in the equation~\eqref{eq:stream_b} corresponds to a weighted reordering of the input $A$ with each kernels and positions assigned to the postsynaptic neurons~\citep{grimaldi_learning_2022}. Let's define \emph{evidence} as the logit of a probability, that is, the inverse sigmoid of that probability. By this definition of evidence, logistic regression takes advantage of the fact that if different independent observations (here, the estimated motion at different spatial locations and timings) share a common cause (here, the rigid local motion of the image on the receptive field), then an optimal estimate of the evidence of this motion is the sum of the evidences from the independent sources. Interpreting the weights of the kernel as evidences (also called factors in logistic regression), we may therefore define the activity $B$ of post-synaptic neurons as the integration of this activity in each voxel and for each channel $\class$ in order to infer the evidence of each motion:
%
\begin{equation}\label{eq:kernel_b}
\forall x, y, \timev, \; \;
B(\class, \timev, x, y)
= \sum_{p, \synapticdelay_\timev, \delta_x, \delta_y} \kernel(\class, p, \synapticdelay_\timev, \delta_x, \delta_y) \cdot A(p, \timev-\synapticdelay_\timev, x - \delta_x, y - \delta_y)
\end{equation}
%
where $\delta_x$ and $\delta_y$ are the relative addresses of the synapses within a kernel and $\delta_\timev$ is the synaptic delay. In this formulation, we recognize that it takes advantage of the position invariance observed in images and exploited in CNNs. Here, we further assume that synaptic motifs should be similar across different times as defined in the temporal convolution. As a consequence, this defines a 3D, spatiotemporal convolutional operator, in which the layers of neurons assigned to specific kernels form channels. Using this dense representation, the model's processing of the input $A$ can be written as layer-wise convolution: $B = \kernel \ast A$ (\seeFig{model} for an illustration). 

The well-known convolution defines a differentiable measure, which is very efficiently implemented for GPUs, and which we will use to detect the motion direction in the event stream. 
A similar type of spatiotemporal filtering was used as a preprocessing stage for an existing pattern recognition algorithm~\citep{ghosh_spatiotemporal_2019}. In addition,~\citet{sekikawa_constant_2018} developed an efficient 3D convolutional algorithm that implements a motion estimation task. By assuming locally a constant motion, the authors assume that the 3D kernel can be decomposed into a 2D kernel representing the shapes and a 3D kernel representing the motion.  For convenience, the connectivity of the neuron $\postsynaddr$ is defined locally around its position $(x_\postsynaddr, y_\postsynaddr)$. Furthermore, it is important to consider that in order to adhere to the limitations of causal computation using biologically realistic neurons, synaptic delays are assigned positive values. This ensures that only past information contributes to the inference made at the present moment. In practice, the kernels are temporally shifted so that the inference at the present time is solely influenced by past information. This temporal shift occurs after a duration equivalent to the depth of the kernel, denoted as $\Ktime$. 

Such a method contrasts with classical methods for delay learning, which explicitly manipulate the delay as a variable and which are not directly differentiable~\citep{nadafian_bio-plausible_2020}. Keeping the analogy with spiking neurons, the analog activity $B$ represents the integration of synaptic activity, and we will now try to define the detection of motion using the spatiotemporal kernels. Since we know that at each instant, there may be different motions, we will define the activation function of our model as a sigmoid function that implements a form of Multinomial Logistic Regression (MLR). In our MLR model, a probability value for each class (i.e., each direction of motion) is predicted for each position $x, y$ and time $\timev$ as a sigmoid function $\sigma(\beta) = \frac {1} {1 + \exp(-\beta)}$ of the result of the convolution. Formally, using the kernels, the input raster plot is transformed into a probability with the following formula:
\begin{equation}\label{eq:mlr}
\forall x, y, \timev, \forall \class \in [1, \Nclass], \; \;
Pr(k=\class \; \vert \; x, y, \timev) =
\sigma (B(\class, \timev, x, y) +\bias_\class) 
\end{equation} 
where $\bias_\class$ is a scalar representing the bias associated with the class $\class$. In particular, we anticipate that certain specific patterns could result in closely synchronized outputs when they are integrated within the basal dendritic tree, consequently leading to heightened postsynaptic activity. By utilizing this analog representation of the evidence for each potential motion at every moment, we can progressively increase the likelihood of generating an output spike. To determine the spiking output, we establish a firing threshold. Here, we computed this threshold to ensure that neurons, on average, generate one spike per second. Therefore, the spiking output of the model corresponds to the motions in space and time that represent the highest probability.

Now that this general framework has been described formally, we may include some heuristics based on neuroscientific observations to constrain our model and its strategies for solving the ecological task described in section~\ref{sec:task}. Note that the general framework is an extension to that presented in~\cite{grimaldi_learning_2022}, in particular by: including a more complex task, the deeper analysis of the results, and these novel neuroscience-inspired heuristics. First, to avoid introducing biases in the directions which may be learned, we apply a circular mask to the spatial dimensions of the kernels. We also included a prior in the selectable motions, as there is a prior for slow speeds in natural scenes~\citep{vacher_bayesian_2018}. Since we want to capture the possible convergence of the trajectories of the events converging on each voxel, we apply a mask to the spatiotemporal kernels such that the smaller the delay, the smaller the radius of the circular mask that is applied (\seeFig{kernels} for an illustration). In our simulations, we observed that including this prior accelerated the learning but was not necessary to reach convergence. 
Second, we observed that moving images produced trajectories of ON and OFF spikes, and that these were present in both polarities. This is due to the fact that our whitened images have a relative symmetry in the luminance profiles, that is, that an image with inverted contrast is indistinguishable from a standard one. Since this arrangement of polarities is independent of motion, we added a mechanism that collects the linear values for the movie and that with the ON and OFF cells flipped, keeping only the maximum value for each voxel. This is similar to the computation done for complex cells in primary visual cortex.

\subsection{Supervised learning of the motion detection task}
Since the model is fully differentiable, we can now implement a supervised learning rule to learn the weights of the model's kernel. This rule was implemented using the binary input events as inputs and the corresponding motion's labels as the desired output. The loss function of the MLR model is the binary cross entropy of the output of the classification layer knowing the ground truth. The labels were defined at each time point as a one-hot encoding of the current motion in the channel corresponding to the current motion, and applied for all positions. Note that in this context, the label is known, but the position of visual features is not, mainly due to the sparse spatial content of natural images. However, the supervised optimization of this MLR model adjusts the weights of the kernels. As a result, we observed that the error is only propagated back to the spatial locations of these most active cells. This is reminiscent of previous methods that solve this problem using a winner-takes-all mechanism~\citep{masquelier_unsupervised_2007}, but is implicit in our formulation. Simulations are performed with the PyTorch library using gradient descent with Adam (for $2^{10}$ movies, each of size $200 \times 128 \times 128$, a learning rate of $10^{-5}$ and $100$ epochs). 

Finally, the output of the MLR model is a representation that predicts the probability of each motion at each position and time. Such an output provides a form of optical flow that can be exploited for non-rigid motion, but we have defined here, for simplicity, an evaluation method that applies to our full-field motion task. 
Using the properties of logistic regression, by taking the mean evidence represented in the output given by the model at all positions for any given time, and using the sigmoid function, we can derive each motion's probability at that time. Taking the most probable class as the output, this allows one to calculate the accuracy as the percentage of times the motion is accurately predicted at any given time step. For validation, these calculations are performed on a different input dataset than the one used in the training or validation steps. The complete code to reproduce the results of this paper is available at \url{https://github.com/SpikeAI/2023_GrimaldiPerrinet_HeterogeneousDelaySNN} (see Data Availability). %
\section{Results}
\label{sec:results}
\subsection{Kernels learned for motion detection}
\begin{figure*}[ht!]
    {\centering
    \includegraphics[width=\linewidth]{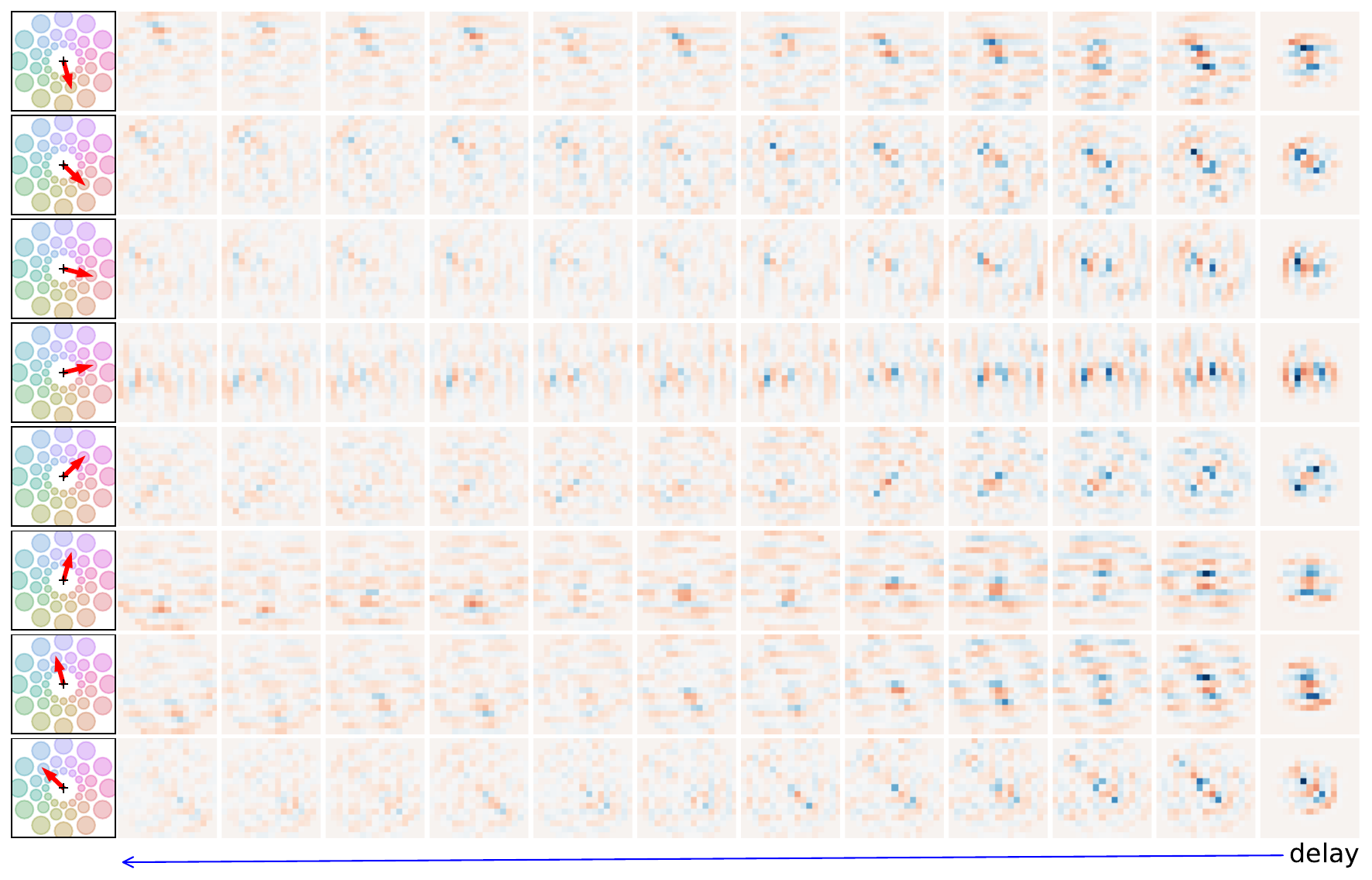}
    }
    \caption{
    	Representation of the weights for $8$ directions for a single speed (among the $12 \times 3$ different kernels of the model) as learned on the dataset of naturalistic scenes. The directions are shown as red arrows in the left insets, where the disks correspond to the set of different possible motions. The spatiotemporal kernels are shown as slices of spatial weights at different delays. Delays vary along the horizontal axis from the far right (delay of one step) to the left (up to a delay of $12$ steps, the remaining synapses being not represented). Each image corresponds to the weights at a given delay, with excitatory and inhibitory weights in warm and cold colors, respectively. Due to the symmetry between the ON and OFF event streams, we observed that the kernels for the OFF polarities are very similar and are not shown here.	Different kernels are selective for the different motion directions, and we observe a slight orientation preference perpendicular to the respective direction for all kernels.
	}
    \label{fig:kernels}
\end{figure*} 
Once our model has been trained, we can begin by examining the learned weights for the various motions (\seeFig{kernels}). Notably, when we track each spatial motif from the shortest delay (on the right) to the longest delay (on the left), we observe that the cells exhibit highly localized selectivity and their preferences are conveyed along linear trajectories in the space-delay domain. When focusing on the positive weights, we notice a pronounced selectivity along specific motion axes for each kernel, and these directions correspond closely to the associated motion's physical direction in visual space. For instance, the first kernel demonstrates a robust preference for downward motion. The negative weights are symmetrically arranged around these positive weights, forming a center-surround profile that is known to enhance the response. We also observed a strong dependence between the weights reaching the ON polarities and those reaching the OFF polarities. In particular, whenever a weight for a given position and delay is positive for one polarity, it will be negative for the other. This property is due to the way events are generated and the fact that the luminance cannot increase and decrease at the same time.  Interestingly, the relative organization of the receptive fields that we observe is in quadrature of phase and follows a push-pull organization predicted by~\citet{kremkow_push-pull_2016} to explain neurophysiological results obtained after showing similar natural scenes with synthetic eye movements~\citep{baudot_animation_2013}. Finally, we observe that these receptive fields show also a relative selectivity to the orientation perpendicular to motion, similarly to what is found for neurons in cortical area MT which is known to be selective to visual motions~\citep{deangelis_functional_1999}. This reflects the way events are generated and in particular the so-called aperture problem which implies that a line moving along its axis would generate no change in luminance and therefore generate no event~\citep{perrinet_motion-based_2012}.

If we now widen our focus on the interpretation of these kernels in terms of spatiotemporal motifs embedded in the event stream, these show a prototypical anisotropic profile adapted to motion detection~\citep{kaplan_anisotropic_2013}. In~\citep{grimaldi_robust_2022}, a link was drawn between event-based MLR training and Hebbian learning, allowing to say that the present model learns its weights according to a presynaptic activity associated to the different motion directions. Each neuron becomes selective to a specific motion direction through the learning of an associated prototypical spatiotemporal spiking motif. Each voxel in the 3D kernels defines a specific property by associating a weight to a position and a delay. Consequently, our model is able to detect precise spatiotemporal motifs embedded in the spike train and associated to the different motion directions. Note that as the delays become larger, two effects can be remarked. First, coefficients become lower which is consistent with the fact that trajectories are defined in a piece wise fashion, such that this decrease provides with an optimal integration considering the gradual diminishing of evidence as time progresses~\citep{Pasturel2020}. Second, coefficients become less localized compared to the kernel's spatial profile at short delays, consistent with the average diffusion of information included in the generative model and with the diffusion introduced in motion-based prediction models~\citep{perrinet_motion-based_2012,khoei_flash-lag_2017}. 
\subsection{Accuracy versus efficiency trade-off}
%
\begin{figure}
    \centering
    \includegraphics[width=0.95\linewidth]{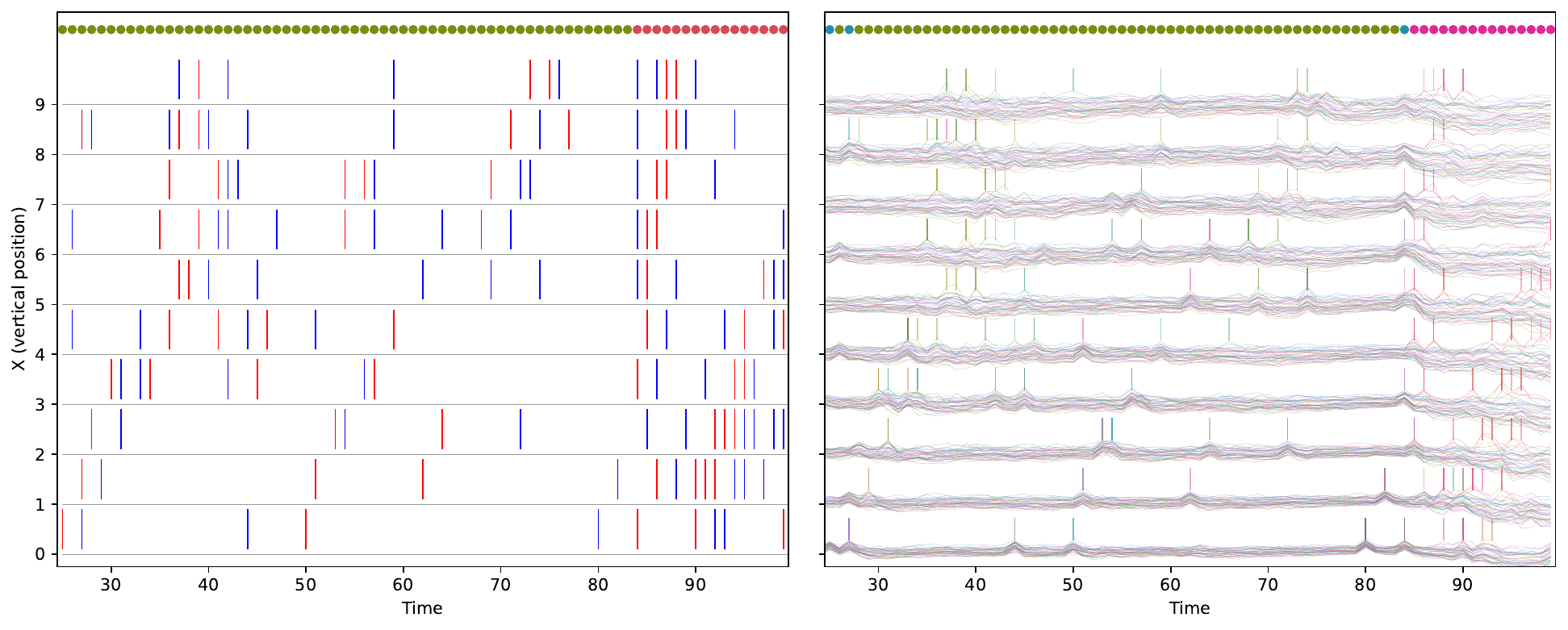}
    \caption{
        In response to a specific event-based input instance (Left), we present the neural activity of the HD-SNN model (Right). To aid visualization, we display a single spatio-temporal slice for a given vertical position ($y = 32$) and $10$ horizontal positions. The input spiking activity comprises ON and OFF spikes, as explained in Fig.~\ref{fig:motion_task}, showcasing the switching within the naturalistic event-based stream from one motion to another due to saccades. The dots above the graph indicate the corresponding motion class at each instant (the motion being represented by the matching color). The output activity consists of two components. Firstly, there is an analog component that corresponds to the evidence accumulated by the model on the spatio-temporal kernels. Secondly, there is a spiking component represented by vertical bars superimposed on the analog activity. These spikes signify moments when the evidence surpasses the spiking threshold. Importantly, this activity aligns with the motion depicted in the input stream. Finally, it is possible to compute the accuracy by comparing the ground truth motion in the input video with the motion predicted by the model (as represented by the colored dots on top of the graph).
        }
    \label{fig:activity}
\end{figure}
After training our MLR model, we obtain spatiotemporal kernels corresponding to the weights associated with the heterogeneous delays of our layer of spiking neurons, which can be used for detection. For this, we quantify its ability to categorize different motions, i.e. on event streams for which the ground truth motion is known at each instant. When applied to new instances of the input movies, the model develops a neural activity which may be used to infer the correct motion~(\seeFig{activity}) and from which we may deduce an accuracy value. This accuracy was computed on a novel dataset of $200$ novel movies. The accuracy computed on the test set was approximately $91\%$ (with a chance level of $1 / 12/ 3 \approx 2\%$). 

\begin{figure}
    \centering
    \includegraphics[width=0.95\linewidth]{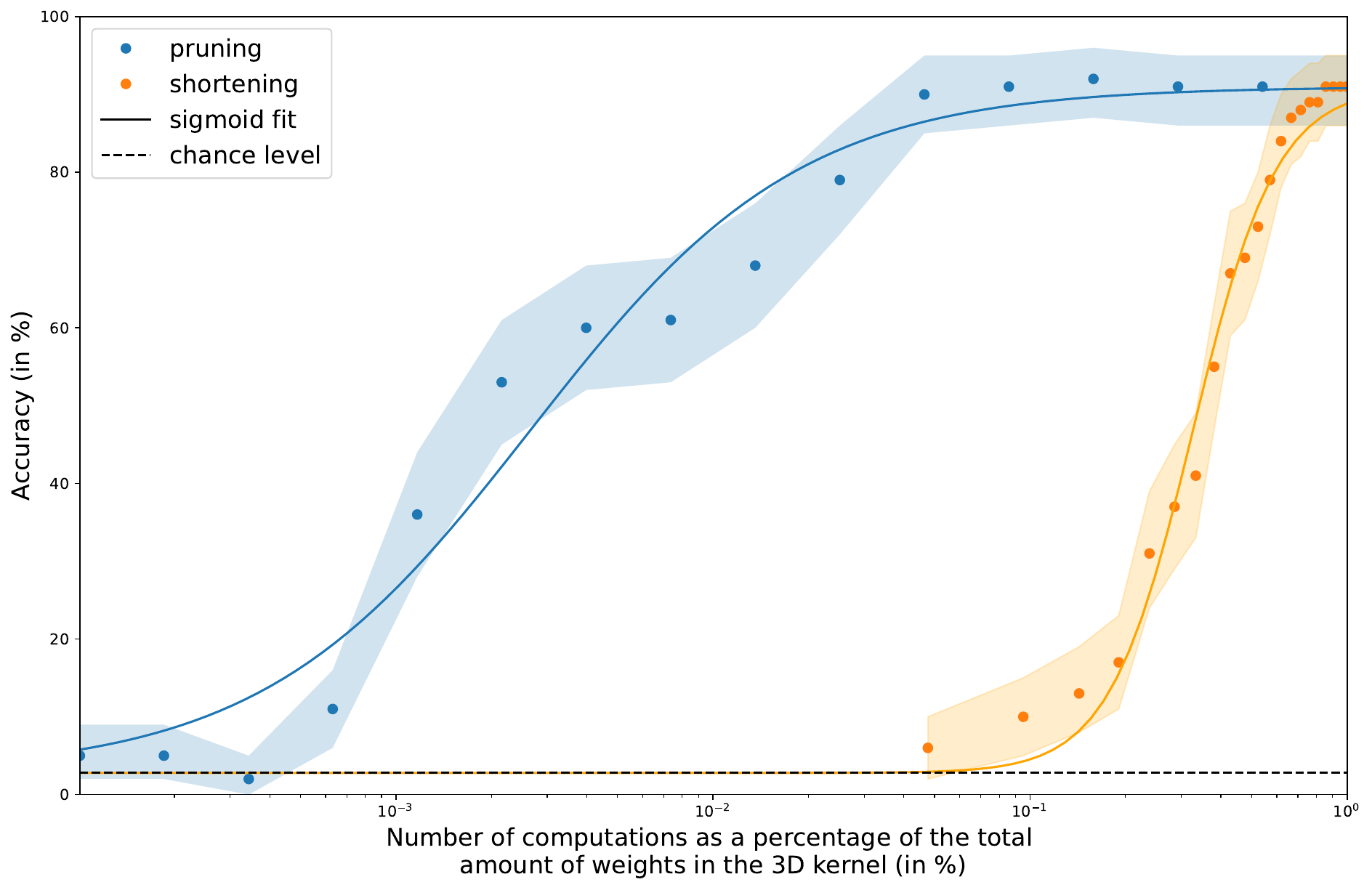}
    \caption{
        Accuracy as a function of computational load for the HD-SNN model (blue dots) with error bars indicating the 5\% - 95\% quantiles and a sigmoid fit (blue line). The relative computational load (on a logarithmic axis) is controlled by changing the percentage of nonzero weights relative to the dense convolution kernel. If we shorten the length of the kernel by using only the weights at the shortest delays, the accuracy quickly drops. However, if we prune the lowest coefficients from the whole kernel, we observe a stable accuracy value, with a drop to half-saturation observed at about $375$ times fewer computations.
        }
    \label{fig:accuracy}
\end{figure}
We also observed that the distribution of the kernel's weights is sparse, with most values close to zero (\seeFig{kernels}). As shown in the formalization of our event-based model, the computational cost of our model, if implemented on a neuromorphic chip, would be dominated by the computations used for the convolution operation. In a dense setting, this corresponds for all voxels in the output to a sum over all voxels in the inputs for all weights in the kernel. But if the information support is sparse, then computations can be now performed only on those events. Specifically, if we set some weights of the kernels to zero, then the additive operation in the convolution for those addresses can be dropped. As a consequence, computations will be performed only on those events which were multiplexed by the pruned connectivity matrix. Thus, knowing the sparseness of the input, the total number of computations scales with the number of spikes multiplied by the number of non-zero synaptic weights. This hypothesis is consistent with biological observations which have shown that communication consumes $35$ times more energy than computation in the human cortex~\citep{Levy2021}.

In order to evaluate the resilience of the classification performance with respect to computational load, we adopt first a pruning approach, where we remove weights in $\kernel$ that fall below a specified threshold. The accuracy of classification is then plotted as a function of the relative number of computations or active weights per decision for each neuron in the layer (refer to Fig.~\ref{fig:accuracy}). To provide a basis for comparison and to account for the benefits of utilizing variable delays, we also present the accuracy achieved by an MLR model employing a shortening strategy. This strategy involves adjusting the temporal width by selecting only the weights associated with the shortest delays. In comparison to the inference performed using the complete 3D kernels without any pruning ($36\times2\times21\times17\times17$), both approaches demonstrate a reduction in computational requirements as indicated by the number of non-zero weights.

By selectively setting certain weights to zero, we observe that the accuracy's evolution, as a function of the logarithmic percentage of active weights, aligns well with a sigmoid curve for both pruning and shortening strategies. The shortening strategy (depicted in orange) demonstrates a rapid decline in accuracy, reaching half-saturation when approximately one-third of the weights remain. On the other hand, the pruning strategy (shown in blue) exhibits a different behavior. It reaches half-saturation when the ratio of active weights is approximately $2.6\times 10^{-3}$, corresponding to around $375$ less computations compared to the dense scenario. In comparison to using the complete kernels, our method maintains accuracy close to its peak performance even when the number of computations is divided by a factor of up to around $31$. This substantial reduction in computations showcases the robustness of the presented method.
%
%
%
\subsection{Testing with natural-like textures}
\begin{figure}
    \centering
    \includegraphics[width=0.99\linewidth]{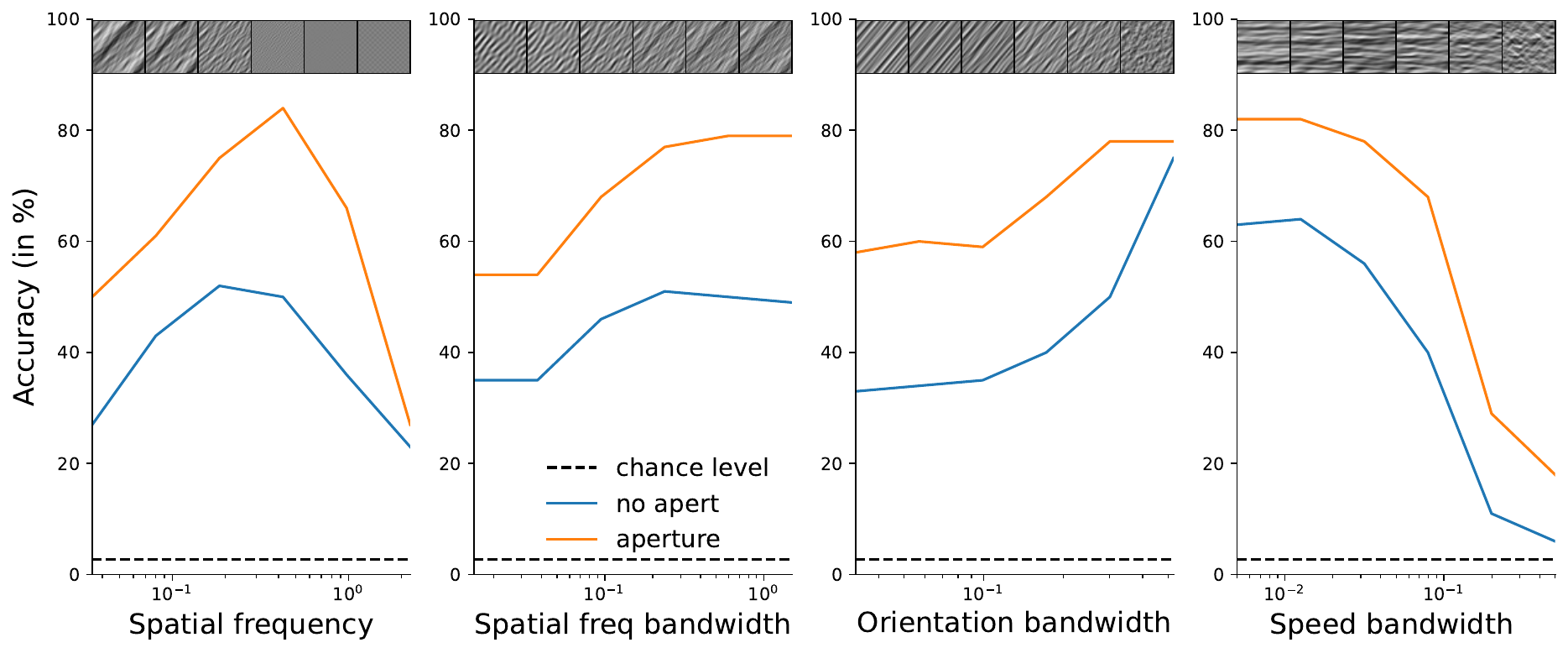} %
    \caption{{\bf Role of stimulus parameters in motion detection accuracy.} Accuracy as a function of (from left to right) mean spatial frequency, bandwidth in spatial frequency (from gratings (left) to isotropic textures (right)), bandwidth in orientation (from isotropic textures (left) to gratings (right)), bandwidth in speed (from a rigid motion (left) to independent frames (right)). Examples snapshots are shown as an illustration in the top insets. Note that these accuracies are computed both in the case where the orientation of the synthetic texture is necessarily perpendicular to the motion ('no aperture' condition) and in the generic case where the orientation is independent of direction ('aperture').
    }
    \label{fig:motion_clouds}
\end{figure}

In order to assess the influence of spatiotemporal parameters of the stimuli on the performance of the model, we now test the model on simpler, parameterized stimuli. For this purpose, we use a set of synthetic visual stimuli, \textit{Motion Clouds}~\citep{leon_motion_2012}, which are natural-like random textures for which we can control relevant parameters for motion detection, including motion direction, spatial orientation, and spatial frequency along with their respective precisions (\seeFig{motion_clouds})~\citep{Leon2012,vacher_bayesian_2018}. By matching the spatial and temporal characteristics of the generated movies with those of the motion task mentioned earlier, we created a range of textures featuring different spatial properties and motions. This procedure defines a set of textures with different spatial properties and different motions chosen from the same set of $12$ directions and $3$ speeds. For each motion, we also varied the texture parameters, such as mean and variance of orientation or spatial frequency content, to provide some naturalistic variability. This method provides a rich dataset of textured movies for which we know the ground truth for the motion.

We observe some interesting facts. First, as we change the mean spatial frequency of the texture, we observe a broadly tuned response in accuracy. This comes as a similar trend as shown in the primary visual areas~\citep{priebe_tuning_2006,Perrinet2007} and reveals the most informative scales learned by our model. Then, by modifying the bandwidth in spatial frequency, we show that the accuracy is worse for a grating-like stimulus than for a large one (which qualitatively resembles a more textured stimulus), reminiscent of the behavioral response of humans to such stimuli~\citep{simoncini_more_2012, ravello_speed-selectivity_2019}. Interestingly, we also see a modulation of accuracy as a function of orientation bandwidth. When the stimulus is grating-like and the orientation is arbitrary with respect to the direction of motion, the system faces the aperture problem and experiences a sharp decrease in accuracy. This is not the case for isotropic stimuli or when the orientation is perpendicular to the direction of motion. Finally, we manipulated the amount of change between two successive frames, similar to a temperature parameter. This shows a progressive decrease in accuracy, similar to that observed in the amplitude of human eye movements~\citep{mansour_pour_speed_2018}.
\section{Discussion}
This paper presents a novel and versatile heterogeneous delay spiking neural network (HD-SNN) that was trained using supervised learning for visual motion detection. We demonstrate the effectiveness of our model by comparing its performance to other event-based classification algorithms for this specific task. Notably, the learned model exhibits similarities with neurobiological and behavioral observations.
One key advantage of our approach is the ability to significantly reduce the computational requirements through synapse pruning, while still maintaining robust classification performance. This highlights the potential to leverage the precise timing of spikes to enhance the efficiency and effectiveness of neural computations.
Overall, our findings underscore the potential of incorporating precise spike timing in neural models and demonstrate the promising capabilities of our heterogeneous delay SNN for event-based computations, specifically in the context of visual motion detection.
\subsection{Synthesis and main contributions}
The HD-SNN model was trained and evaluated on a naturalistic motion detection task with realistic eye movements. It is defined such as to provide an optimal detection of  spatiotemporal motifs and learns kernels similar to those found in the visual cortex~\cite{deangelis_functional_1999, kremkow_push-pull_2016}. We have evaluated the computational cost of this model when implemented in a setting similar to event-based hardware. We show that the use of heterogeneous delays may be an efficient computational solution for future neuromorphic hardware, but also a key to understanding why spikes are a universal component of neural information processing.

We would like to highlight a few innovations in the contributions presented in this paper. First, while~\citep{ghosh_spatiotemporal_2019,yu_stsc-snn_2022} use a correlation-based heuristic, the generic heterogeneous model is formalized from first principles for optimal detection of spatiotemporal spiking motifs using a time-invariant logistic regression. Moreover, compared to classical CNN solutions, the parameters of this one-layered model (weights and delays) are explainable, as they directly inform about the evidence  of detection for each spatiotemporal spike motif, where we define \emph{evidence} as the logit of the probability, that is, the inverse sigmoid of the probability. Another novelty is that the model learns simultaneously weights and delays. In contrast, the polychronization model~\citep{izhikevich_polychronization_2006} learns only the weights using STDP, while the delays are randomly drawn at initialization and their values are frozen during learning. In addition, the model is evaluated on a realistic task, while models such as the tempotron are tested on simplified toy problems~\citep{gutig_tempotron_2006}. Another major contribution is to provide a model that is suitable for learning any kind of spatiotemporal spiking motif and that can be trained in a supervised manner by providing a dataset of supervised pairs. Instead of relying on a careful description of the physical rules governing a task, e.g. the luminance conservation principle for motion detection~\citep{benosman_asynchronous_2012, dardelet_event-by-event_2021}, this allows a more flexible definition of the model using this properly labeled dataset.
\subsection{Main limits} 
We have identified a number of limitations of our model, which we will now discuss in detail. First, this implementation of the HD-SNN model is based on a discrete binning of time, which is not compatible with the continuous nature of biological time. We used this binning to efficiently implement the framework on conventional hardware, especially GPUs, in particular to be able to use fast, differentiable three-dimensional convolutions. This is consistent with the relative robustness of other event-based frameworks~\citep{lagorce_hots_2017,grimaldi_robust_2022}, where accuracy was unaffected when the input spikes were subjected to noisy perturbations up to $4~\ms$ on the N-MNIST dataset~\citep{grimaldi_robust_2022}. It suggests the potential advantage of analytically including an additional precision term to the temporal value of input spikes. Such a mechanism may be implemented by the filtering implemented by the synaptic time constant of about $5~\ms$. Furthermore, it is possible to circumvent the need for time discretization by the use of a purely event-based scheme. In fact, it is possible to derive event-triggered computations of the continuous activity of the SNN~\citep{hanuschkin_general_2010} and thus to define a purely event-based framework. Such an architecture could provide promising computational speedups.

Another limitation is that the model is purely feed-forward. Thus, the spikes generated by the postsynaptic neurons are based solely on the information contained in the classical receptive field. However, it is known that neurons in the same layer can interact with each other through lateral interactions, for example in V1, and that this can be the basis for more complex computational principles~\citep{chavane_revisiting_2022}. For example, the combination of neighboring orientations may contribute to image categorization~\citep{perrinet_edge_2015}. Furthermore, neural information may be modulated by feedback information, e.g. to distinguish a figure from its background~\citep{roelfsema_early_2016}. Feedback has been shown to be essential for building realistic models of primary visual areas~\citep{boutin_sparse_2020, boutin_effect_2020}, especially to explain non-linear mechanisms~\citep{boutin_pooling_2022}. Currently, mainly due to our use of convolutions, it is not possible to implement these recurrent connections in our implementation (lateral or feedback). However, by inserting new spikes into the list of spikes reaching presynaptic addresses, the generic HD-SNN model is able to incorporate them. While this is theoretically possible, it must be properly tuned in practice so that these recurrent connections do not bring neuronal activity outside a homeostatic state (by extinction or explosion).

Such recurrent activity would be essential for the implementation of predictive or anticipatory processes~\citep{Benvenuti2020}. This is essential in a neural system because it contains several delays that require temporal alignment~\citep{hogendoorn_predictive_2019}. This has been modeled before to explain, for example, the flash-lag illusion~\citep{khoei_flash-lag_2017}. As mentioned previously, this could be implemented using generalized coordinates (i.e., variables such as position complemented by motion, acceleration, jerk,~\ldots), and  knowing that ``neurobiologically, using delay operators just means changing synaptic connection strengths to take different mixtures of generalized sensations and their prediction errors''~\citep{perrinet_active_2014}. Our proposed model using heterogeneous delays provides an alternative and elegant implementation solution to this problem.
\subsection{Perspectives}
%
In defining our task, we emphasized that the generation of events depends on the spatial gradient in each image. This gradient has both horizontal and vertical dimensions, and its maxima are generally orientation dependent. Taken together, these oriented edges form the contours of visual objects in the scene~\citep{koenderink_representation_1987}. Thus, there is an interdependence between motion information and orientation information within the event stream, which we put in evidence by the shape of the kernels. It would be crucial to investigate this dependency further. This could be initiated by training the model on a dataset with labels that provide local orientation. Exploring this dependence will allow us to dissociate and integrate these two forms of visual information. In particular, it will allow us to consider that the definition of motion is more accurate perpendicular to an oriented contour (that is, the aperture problem). Thus, it will allow us to implement recurrent prediction rules, such as those identified to dissociate this problem~\citep{perrinet_motion-based_2012}.

The model is trained on a low-level local motion detection task, and one might wonder if it could be trained on higher-level tasks. An example of such a task would be depth estimation in the visual scene. There are several sources of information for depth estimation, such as binocular disparity or changes in texture or shading, but in our case motion parallax would be the most important cue~\citep{rogers_motion_1979}. This is because objects that are close to an observer move on the retina relatively faster than an object that is far away, and also because visual occlusions depend on depth order. Using this information, it is possible to segment objects and estimate their depth~\citep{yoonessi_contribution_2011}. However, this would first require the computation of the optic flow, i.e., the extension of the framework described here for a rigid full-field motion to a more general one where the motion may vary in the visual field. One possible implementation is to add a new layer to our model, analogous to the hierarchical organization which is prevalent in the visual cortex. This is theoretically possible by using the output of our model (which estimates motion in retinotopic space) as input to a new layer of neurons that would estimate motion in the visual field, including the depth dimension in the output supervision labels. This could have direct and important applications, e.g. in autonomous driving to detect obstacles in a fast and robust way. Another extension would be to actively generate sensor motion (physical or virtual) to obtain better depth estimates, especially to disambiguate uncertain estimates~\citep{nawrot_eye_2003}.

In conclusion, the HD-SNN model that we have presented provides a way to efficiently process event-based signals. We have shown that we can train the model using a supervised rule that infers \emph{what} is the output label, but not \emph{where} it occurs. Another perspective would be to extend the model to a fully self-supervised learning paradigm, i.e. without any labelled data~\citep{barlow_unsupervised_1989}. This type of learning is thought to be prevalent in the central nervous system and, assuming the signal is sparse~\citep{olshausen_emergence_1996}, one could extend these Hebbian sparse learning schemes to spikes~\citep{perrinet_emergence_2004, masquelier_competitive_2009}. We expect that this would be particularly useful for exploring neurobiological data~\citep{KeatingPerrinet23ICANN}. Indeed, there is a large literature showing that brain dynamics often organize into stereotyped sequences, such as synfire chains~\citep{ikegaya_synfire_2004}, packets~\citep{luczak_sequential_2007}, or hippocampal sequences~\citep{pastalkova_internally_2008, villette_internally_2015}. These motifs are stereotyped and robust, as they can be activated following the same motif from day to day~\citep{haimerl_internal_2019}. In contrast to conventional methods of processing neurobiological data, such an event-based model would be able to answer key questions about the representation of information in neurobiological data, and it would open possibilities in the field of computational neuroscience. Furthermore, it would open possibilities in the field of machine learning, especially in computer vision, to address current key concerns such as robustness to attacks, scalability, interpretability, or energy consumption.
%
\backmatter
\bmhead{Acknowledgments}
\Acknowledgments
%

\section*{Statements and Declarations}

\paragraph{Funding}

\Funding %

\paragraph{Conflict of interest}
Not applicable.

\paragraph{Ethics approval}
Not applicable.

\paragraph{Consent to participate}
Not applicable.

\paragraph{Consent for publication}
Not applicable.

\paragraph{Availability of data and materials}
Not applicable.

\paragraph{Code availability}
\DataAvailability
\paragraph{Authors' contributions}
Both authors contributed to the conceptualization and methodology design of the study, to the project's coordination and administration. Laurent Perrinet carried out the funding acquisition and supervision. Formal analysis and investigation were performed by both authors. Results visualization and presentation were realized by both authors. The manuscript was written by both authors. Both authors have read and approved the final manuscript.
\bibliographystyle{apalike}
\bibliography{FastMotionDetection}
\end{document}